\documentstyle{elsart}

\begin{document}
 
\begin{frontmatter}
\title{A practical criterion of irreducibility of multi--loop Feynman 
integrals.
}


\author{P.\,A.\,BAIKOV \thanksref{AA}}

\thanks[AA]{Supported in part by INTAS (grant 03-51-4007)
and RFBR (grant 05-02-17645);
e-mail: baikov@theory.sinp.msu.ru}

\address{Skobeltsyn Institute of Nuclear Physics, Moscow State 
University,\\
Moscow~119992, Russia}

\begin{abstract}
A practical criterion for the irreducibility
(with respect to integration by part identities)
of a particular Feynman integral to a given set of integrals
is presented.
The irreducibility is shown to be related to the existence of stable
(with zero gradient) points of a specially constructed polynomial.
\end{abstract}

\end{frontmatter}

The growing accuracy of high energy physics experiments
demands the calculation of higher order quantum  corrections.
The latter, in turn, are expressed through multi-loop
Feynman integrals.
Most of the methods for their calculations 
are designed for the integrals with a monomial of momenta in the numerator. 
On the other hand, in practice
one can face
thousands and even millions of such integrals in one particular 
physical problem.
Fortunately, it is possible to reduce this number to around a
few dozens by using integration by part identities \cite{ch-tk}.
These identities can be solved 
in some cases by constructing recursive algorithms \cite{ch-tk,alg}
or by "brute computer force" with the Laporta algorithm \cite{Laporta}.
In any case, it is useful to know if the given integral can be related
to simpler integrals (with some denominators missing).
A corresponding criterion of irreducibility was proposed in \cite{oldcrit}:
the irreducibility was related to the existence of a special
solution of the recurrence relations; a recipe to construct such solutions 
was suggested. 
Unfortunately,  the recipe in the form of \cite{oldcrit} required some "hand work" which
becomes very undesirable for the real--life problems 
(too many cases to be considered). So in this paper we propose a
more practical version of this criterion which amounts to the study of
zero-gradient points of some polynomial and hence can be easily implemented with
a computer.
We start with examples, then formulate the criterion and finally present the list
of the irreducible four-loop massless propagator integrals found with the help of our 
criterion.

\section{Examples.}

Consider two massless 4-loop propagator integrals:

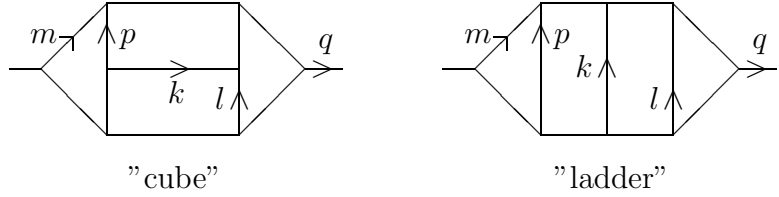
\begin{center} 
\begin{picture}(125,50) 
\put(45,-20){"cube"}
\put(12,25){\line( -1, 0){12}}
\put(12,25){\line( 1, 1){25}}  \put(24,37){\line(-1,0){5}}\put(24,37){\line(0,-1){5}}
\put(8,35){$m$}
\put(12,25){\line( 1,-1){25}}

\put(87,50){\line( 0, -1){50}} \put(83,10){$\wedge$} \put(78,10){$l$}
\put(37,50){\line( 1,  0){50}}
\put(37,  0){\line( 0,  1){50}} \put(33,35){$\wedge$} \put(42,35){$p$}
\put(37,  0){\line( 1,  0){50}}

\put(112,25){\line(-1,-1){25}} 
\put(112,25){\line(-1, 1){25}}
\put(112,25){\line( 1, 0){14}} \put(117,33){$q$}
\put(114,22){$>$}

\put(37,25){\line( 1, 0){50}}\put(60,22){$>$}\put(60,12){$k$}
\end{picture}
\hskip 3em
\begin{picture}(125,50) 

\put(42,-20){"ladder"}

\put(12,25){\line( -1, 0){12}}
\put(12,25){\line( 1, 1){25}}  
\put(24,37){\line(-1,0){5}}\put(24,37){\line(0,-1){5}}
\put(8,35){$m$}
\put(12,25){\line( 1,-1){25}}

\put(87,50){\line( 0, -1){50}} \put(83,10){$\wedge$} \put(78,10){$l$}
\put(37,50){\line( 1,  0){50}} 
\put(37,  0){\line( 0,  1){50}} \put(33,35){$\wedge$} \put(42,35){$p$}
\put(37,  0){\line( 1,  0){50}} 

\put(112,25){\line(-1,-1){25}} 
\put(112,25){\line(-1, 1){25}}
\put(112,25){\line( 1, 0){14}} \put(117,33){$q$}
\put(114,22){$>$}

\put(62,0){\line( 0, 1){50}}\put(58,22){$\wedge$}\put(50,22){$k$}
\end{picture}
\end{center}

\vskip 1em

The "ladder" integral consist of triangles and thus, 
according to "triangle rule" \cite{ch-tk},
can be related to the combination of simpler integrals; the reducibility of the 
"cube" diagram is unclear.
To study this question we are going to associate with every multi--loop 
integral some polynomial, whose properties reflect its irreducibility.

Let us calculate the determinant $P=\det(p_i p_k)$ $(p_i=q,m,p,k,l)$ 
under the condition that all lines are "on mass--shell", that is corresponding 
squared momentum are equal to zero. 
In total there are 15 scalar products ("external" $q^2$ and 14 "internal" involving 
loop momenta) with 11 constraints, so we can express  11 "internal" scalar products 
through others. For "cube" it is convenient to  choose $pq, lp, lq, q^2$  as independent:

$P_{cub} =  - (2\ lp\cdot lq + 2\ lp\cdot pq + lp\cdot q^2 - 2\ lq\cdot pq\ )
\cdot lp\cdot lq\cdot pq.$

For "ladder" let us choose $kl, kp, lp, q^2$ as independent:

$P_{lad} = (kl\cdot kp - lp\cdot q^2/2)\cdot kl\cdot kp.$

For each polynomial let us check if there are points where all derivatives with 
respect to the "internal" scalar products are zero, but the polynomial 
itself and the determinant of second derivatives are not zero 
(in short "non--zero stable points").
The "ladder" polynomial has no such points, and the "cube" has one point:
{$lp=-9q^2/20, pq=-3q^2/10, lq=-3q^2/10$}.
As we will see later, this means irreducibility of the "cube" integral.

\section{Criterion in general form.}

Now let us explain why non-zero stable points of some polynomial are related to
the irreducibility. 
Assume that there is a set of equations which relate
the function $B(\underline{n})$ 
(keeping in mind Feynman integrals as functions of degrees of propagators and 
irreducible numerators) with arguments shifted by $\pm1$:
\begin{equation}
R(I^-,I^+)B(\underline{n})=0,
\label{eqrr}
\end{equation}
where ($\underline{n})$ (and others underlined 
arguments) is $(n_1,..,n_N)$,
${\bf I}^-_a B(.., n_a,..)\equiv B(..,n_a-1,..)$
and ${\bf I}^+_a B(.., n_a,.. )\equiv n_a B(.., n_a+1,..)$.
Since ${\bf I}^-$ and ${\bf I}^+$ do not commute 
($[{\bf I}^-_a,{\bf I}^+_b]=\delta_{ab}$),
for definiteness we place all ${\bf I}^+$ to the right of ${\bf I}^-$.
 
We want to check the possibility to represent, using the relations
(\ref{eqrr}), a given $B(\underline{n}^{(0)})$ as linear
combination of $\{ B(\underline{n}^{(i)}),
i\in(1,...,k)\}$:
\begin{equation}
0 \stackrel{?}{=} B(\underline{n}^{(0)}) - \sum_{i=1}^k f_i B(\underline{n}^{(i)}).
\label{eqstar}
\end{equation}
The relation (\ref{eqstar}), if it exists, should be a combination of 
the relations (\ref{eqrr}) for various values of $\underline{n}$, so
any solution of (\ref{eqrr}) should also obey (\ref{eqstar}).
Suppose we are able to construct the special solution of (\ref{eqrr}) $s(\underline{n})$ 
with the properties
\begin{equation}
s(\underline{n}^{(0)})\neq0, \qquad 
s(\underline{n}^{(i)})=0\quad \mbox{for}\ i\in(1,...,k).
\label{eqnorm}
\end{equation}
This $s(\underline{n})$ evidently cannot obey (\ref{eqstar}) and hence
(\ref{eqstar})
cannot be the combination of (\ref{eqrr}). 
So we got a sufficient criterion
of irreducibility in general form \cite{oldcrit}: 
the recurrence relations (\ref{eqrr}) cannot relate
the $B(\underline{n}^{(0)})$ to
linear combination of the set \{$B(\underline{n}^{(i)})$,
$i\in(1,...,k)$\} if a partial solution of (\ref{eqrr}) with
properties (\ref{eqnorm}) exits.

\section{Formal solutions of the recurrence relations.}

Let us construct the solutions of the relations (\ref{eqrr}) as formal series in $1/D$:
\begin{equation}
B(\underline{n},D)=\frac{1}{D^{k_0(\underline{n})}}\sum_{k=0}^{\infty}\frac{B^{(k)}(\underline{n})}{D^k}.
\label{fserias}
\end{equation}
To be precise, we will present an algorithm for calculation of the $B^{(k)}$ such that
for any $\underline{n}$ and arbitrary large $N$,
sufficiently long partial sum of (\ref{fserias})
will obey (\ref{eqrr}) up to terms $1/D^N$.

As an example, let us consider the recurrence relations
\begin{equation}
0
=(k-2D+2)B_{k+1}+2a(k-D+1)B_k+kbB_{k-1}
\label{recrelex}
\end{equation}
generated by the integral 
\begin{equation}
B_k=\int_{-\infty}^{\infty} x^k (x^2+2ax+b)^{-D} dx, \quad k\geq 0,\ k+1-2D<0, \ b>a^2.
\label{intex}
\end{equation}
After the substitution $x=\tilde{x}/\sqrt{D}-a$ integral (\ref{intex}) can be represented as
\begin{eqnarray}
B_k
&=&\sum_{l=0}^{k}\frac{\Gamma(k+1)}{\Gamma(l+1)\Gamma(k-l+1)}(-a)^{k-l}D^{-l/2-1/2}\tilde{B}_l,
\label{BkBl}\\
\tilde{B}_k&=&\int_{-\infty}^{\infty} \tilde{x}^k (\frac{\tilde{x}^2}{D}+b-a^2)^{-D} d\tilde{x}.
\nonumber
\end{eqnarray}
The $\tilde{B}_k$ obeys the recurrence relations
\begin{equation}
0=\big(k(b-a^2)\tilde{B}_{k-1}-2\tilde{B}_{k+1})+\frac{1}{D}(k+2)\tilde{B}_{k+1}=
R_k^0(\tilde{B})+\frac{1}{D}R_k^1(\tilde{B}).\label{recrelexR}
\end{equation}

Inserting $\tilde{B}_k=\sum_{l=0}^{\infty}\tilde{B}_k^{(l)}/D^l$ we obtain
the following equations for $\tilde{B}_k^{(l)}$:

$$0=R_k^0(\tilde{B}^{(0)}), \quad 0=R_k^0(\tilde{B}^{(l+1)})+R_k^1(\tilde{B}^{(l)}) \qquad l>0.$$

If $b-a^2\neq0$, these equations have nontrivial solution and we obtain the desired solution of the 
(\ref{recrelex}) as formal series in $1/D$.
Note that we can forget about integral (\ref{intex}) and study the relation (\ref{recrelex})
with the help of substitution (\ref{BkBl}). In particular if $b\leq a^2$ the integrand of (\ref{intex}) 
has poles on the real axes, but the substitution (\ref{BkBl}) does not depend on the definition (and 
even on the convergence) of the integral.
From other side, the "integral representation" language is more convenient, so in general 
case we will construct solutions (\ref{eqnorm}) presenting necessary manipulations in this way,
keeping in mind that for formal proofs we can translate all steps to "recurrence relations" language.

\section{Formal solutions in general case.}

To construct the integral representation for $s(\underline{n})$ (\ref{eqnorm}) let us try\cite{me}:

\begin{equation}
s(\underline{n})=\int \frac{dx_1...dx_N}{x_1^{n_1}...
x_N^{n_N}}g(\underline{x}).
\label{eqsol}
\end{equation}
The action of $R(I^-,I^+)$ on (\ref{eqsol}) leads to 
\begin{eqnarray}
R(I^-,I^+) s(\underline{n})&=&\int dx_1...dx_N g(\underline{x})
R(\underline{x},-\partial_{\underline{x}})\frac{1}{x_1^{n_1}...x_N^{n_N}}\nonumber\\
&=&\int\frac{dx_1...dx_N}{x_1^{n_1}... x_N^{n_N}} R'(\underline{x},\partial_{\underline{x}}) 
g(\underline{x}) + (\mbox{surface terms}),
\label{eqrrsol}
\end{eqnarray}
where we use integration by part in second equation 
and denote by $R'$ the combination of $\underline{x}$ and 
$\partial_{\underline{x}}$ acting on $g(\underline{x})$.
Now, if one chooses $g(\underline{x})$ so that
\begin{equation}
R'(\underline{x},\partial_{\underline{x}})g(\underline{x})=0 
\label{eqrrg}
\end{equation}
and chooses the integration contours such that the surface terms vanish,
 one arrives to the function $s(\underline{n})$
such that $R(I^+,I^-)s(\underline{n})=0$.

As shown in \cite{me}, the (\ref{eqrrg}) can be solved for the general
case of a multi-loop Feynman integral with arbitrary number of legs and with
arbitrary masses,
and the corresponding $g(\underline{x})$ can be represented as a product of 
two
polynomials in $\underline{x}$, each polynomial raised to non-integer 
degree.
Then, the recurrence relations for $L$--loop $R$--leg integrals is a partial case of 
the recurrence relations for $R+L-1$ tadpole integrals \cite{equiv}, thus we limit 
ourselves by the tadpole case.

Consider an $L$-loop tadpole Feynman integrals in dimensional regularization:

$$B(\underline{n})\equiv \int \frac{d^Dp_1\ldots d^Dp_L}{D_1^{n_1}\ldots D_N^{n_N}},$$
\nonumber

where $p_i$ (${i=1,\ldots,L}$) are loop momenta and
\begin{equation} 
D_a=\sum_{i=1}^{L}\sum_{j=1}^{L}
A^{ij}_a p_i\cdot p_j - m_a^2 \quad (a=1,\ldots,N=L(L+1)/2) 
\label{defD}
\end{equation}
are propagators ($n_a>0$) or irreducible numerators ($n_a\leq 0$).
Consider some integral $B(\underline{n}^{(0)})$ with
the set of positive indexes ("lines") $S$: $a\in S$ if $n_a^{(0)}>0$.
Normally, the integrals with less number of lines are easier to calculate,
so let us check if it is possible to relate this integral
to integrals with some lines missing. 
For that we should try to construct $s(\underline{n})$ such that $s(\underline{n})=0$ if 
at least one of the $n_{a\in S}\leq 0$.
Eq. (\ref{eqsol}) in this case  will read \cite{me}
\begin{equation}
s(\underline{n})=\int \frac{dx_1...dx_N}{x_1^{n_1}...
x_N^{n_N}}P(\underline{x})^{(D-L-1)/2},
\label{eqsol1}
\end{equation}   
where
$P(\underline{x})=\det_{(kl)}(\sum_{a=1}^{N}\tilde{A}^{kl}_a(x_a+m^2_a))$ 
is polynomial in $\underline{x}$
and $\tilde{A}^{kl}_a$ is the matrix inverse of $A^{kl}_a$, i.e.
$\sum_{a=1}^{N}\tilde{A}^{ij}_a A^{kl}_a=(\delta_{ik} \delta_{jl}+\delta_{il} 
\delta_{jk})/2$.
In other words, to construct $P$ one should calculate the determinant of scalar 
products of the loop momenta and then substitute the scalar products by the combinations
of propagators and numerators obtained by inverting (\ref{defD}).

Let us choose the integration contours for $x_{a\in S}$ as
small circles around zero. In this case, according to Cauchy's theorem,
the integrations will lead to the $(n_a-1)^{th}$ coefficient in
the Taylor expansion of the integrand:
\begin{equation}
s(\underline{n})\propto
\int \prod_{a \notin S}
\left[\frac{dx_a}{x_a^{n_a}}\right]
\prod_{a \in S}\left[\frac{\partial_a^{n_a-1}}
{(n_a-1)!}\right]
P(\underline{x})^{(D-L-1)/2}
\Big|_{x_{a \in S}=0}.
\label{expan}
\end{equation}

In the case when the number of lines is equal to the number of integrations in 
(\ref{eqsol1}) the (\ref{expan}) has no integrations and 
(if $P(\underline{x})\big|_{x_{a \in S}=0}\neq0$) the construction of 
the s(\underline{n})
is completed; in the general case, we need a way to define the integration over the remaining 
$x_{a \notin S}$. As the result of the Taylor expansion, the original $s(\underline{n})$ 
will be linear combination of the terms of the following type: 
\begin{eqnarray}
s(\underline{n})&=& \sum_{k_{a \notin S}} C_{k_{a \notin S}}
\int \prod_{a \notin S}\left[ dx_a x_a^{-k_a}\right]
\tilde{P}(\underline{x})^{D'/2},\nonumber \\
\tilde{P}(\underline{x})&=&P(\underline{x})\big|_{x_{a \in S}=0},\quad D'=D-L-1-\sum_{a 
\in S}(n_a-1).
\label{1D}
\end{eqnarray}
Let us expand (\ref{1D}) in the formal series in the limit $D'\rightarrow -\infty$.
Assume there exists a point $\underline{x}^{0}$ with:
\begin{equation}
\tilde{P}(\underline{x}^{0})\neq0,\quad
\partial_{a \notin S} \tilde{P}(\underline{x}^{0})=0,\quad 
\det_{ab}(\partial_{a \notin S} \partial_{b \notin 
S}\tilde{P}(\underline{x}^{0}))\neq0. 
\label{crit}
\end{equation}
After the substitution $\underline{x}=\underline{x}^{0}+\underline{x}'/\sqrt{-D'}$
\begin{eqnarray}
\tilde{P}(\underline{x})^{\frac{D'}{2}}&\approx&
\Big(\tilde{P}(\underline{x}^{0})
-\sum_{a,b\notin S}\frac{x'_ax'_b}{2D'}\partial_a\partial_b\tilde{P}(\underline{x}^{0})
+O((-D')^{-3/2})\Big)^{\frac{D'}{2}}\nonumber\\
&\approx&\tilde{P}(\underline{x}^{0})^{\frac{D'}{2}}
\Big(e^{-\frac{1}{4\tilde{P}(\underline{x}^{0})}\sum_{a,b\notin 
S}{x'_ax'_b}\partial_a\partial_b\tilde{P}(\underline{x}^{0})}
+O((-D')^{-1/2})
\Big).\label{Pexp}
\end{eqnarray}

The Gaussian integrations can be performed by the formula
\begin{equation}
\int \prod_{a=1}^M (dx_a x_a^{k_a})e^{(-
x_aC^{ab}x_b)}=\frac{\pi^{M/2}}{\det(C)^{1/2}}\prod_{a=1}^M(\frac{d}{dt_a})^{k_a}e^{(
\frac{1}{4}
t_aC^{-1,ab}t_b)}\Big|_{t_a=0}.\label{Gauss}
\end{equation}
Finally, expansion in $D'\rightarrow -\infty$ can be easily recalculated to $D\rightarrow -\infty$. 

Eq. (\ref{crit}) defines the practical version of the criterion:
if (\ref{crit}) is valid, then the eqs. (\ref{expan},\ref{1D},\ref{Pexp},\ref{Gauss}) define the 
algorithm for calculating
$s(\underline{n})$ with properties (\ref{eqnorm})
thus proving the irreducibility of the integral $B(\underline{n}^{(0)})$.
 Note that $P$ in (\ref{eqsol1}) is $\det(p_i p_k)$ where scalar products 
are expressed through linear combinations of $\underline{x}$, and hence $\tilde{P}$ in 
(\ref{1D}) is equivalent to $\det(p_i p_k)$ with lines "on-shell", as it was considered 
in the examples.

The equation ({\ref{crit}}) and hence (\ref{1D}) can have several solutions 
$s^{(p)}(\underline{n})$. This corresponds to possibility to find several 
integrals with the same set of lines (differ by number of dots or by numerators) 
which are irreducible as to each other, as to simpler integrals.
To fix these irreducible integrals,
let us choose some set of points in "$\underline{n}$"-space {$\underline{n}^{(q)}$} such 
that the matrix $M^p_q=s^{(p)}(\underline{n}^{(q)})$ has inverse matrix 
$\overline{M}^q_p$. 
Then solutions $\overline{M}^q_p s^{(p)}$ will be "diagonal" in the sense 
(\ref{eqnorm}) and hence the set {$\underline{n}^{(q)}$} will define 
irreducible integrals.

The application of the criterion (\ref{crit}) to the four--loop massless propagator 
integrals
(or equivalently five--loop tadpole with one massive line) results in the following 
list of integrals which cannot be reduced to integrals with a smaller number of lines
(for each of the topologies depicted here only a single irreducible integral exists):

\vskip -2em
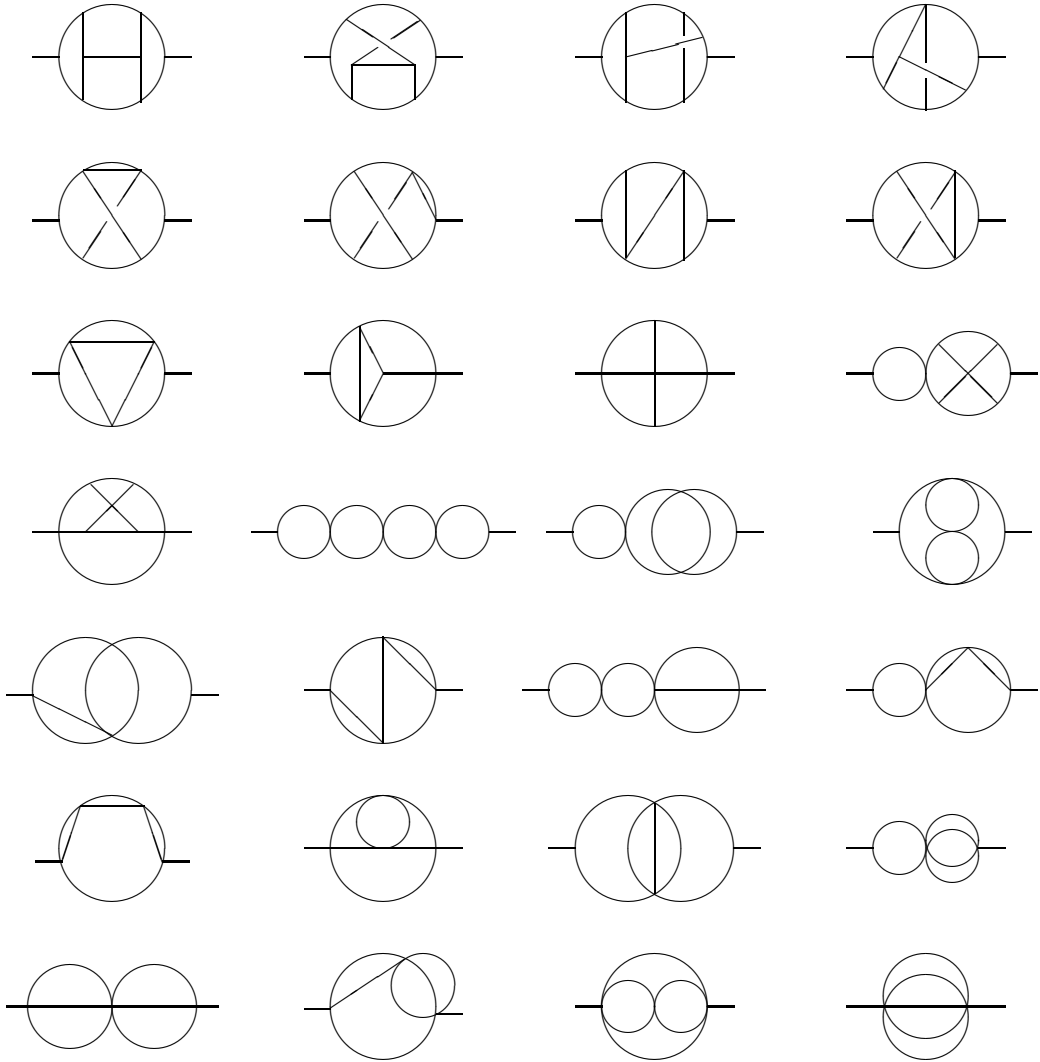
\begin{picture}(100,100) 
\put(50,50){\circle{40}}
\put(70,50){\line(1,0){10}}
\put(30,50){\line(-1,0){10}}
\put(61,67){\line(0,-1){34}}
\put(39,67){\line(0,-1){33}}
\put(39,50){\line(1,0){22}}
\end{picture}
\begin{picture}(100,100) 
\put(50,50){\circle{40}}
\put(70,50){\line(1,0){10}}
\put(30,50){\line(-1,0){10}}
\put(62,34){\line(0,1){13}}
\put(38,34){\line(0,1){13}}
\put(38,47){\line(1,0){24}}
\put(62,47){\line(-3,2){26}}
\put(38,47){\line(3,2){10}}
\put(64,64){\line(-3,-2){11}}
\end{picture}
\begin{picture}(100,100) 
\put(50,50){\circle{40}}
\put(70,50){\line(1,0){10}}
\put(30,50){\line(-1,0){10}}
\put(39,67){\line(0,-1){34}}
\put(39,50){\line(4,1){29}}
\put(61,33){\line(0,1){20}}
\put(61,67){\line(0,-1){9}}
\end{picture}
\begin{picture}(100,100) 
\put(50,50){\circle{40}}
\put(70,50){\line(1,0){10}}
\put(30,50){\line(-1,0){10}}
\put(50,70){\line(-1,-2){16}}
\put(40,50){\line(2,-1){25}}
\put(50,70){\line(0,-1){22}}
\put(50,30){\line(0,1){12}}
\end{picture}
\vskip -5em
\begin{picture}(100,100)  
\put(50,50){\circle{40}}
\put(61,67){\line(-2,-3){9}}
\put(39,34){\line(2,3){9}}
\put(50,50){\line(-2,3){11}}
\put(50,50){\line(2,-3){11}}
\put(61,67){\line(-1,0){22}}
\put(70,48){\line(1,0){10}}
\put(30,48){\line(-1,0){10}}
\end{picture}
\begin{picture}(100,100)  
\put(50,50){\circle{40}}
\put(61,66){\line(-2,-3){9}}
\put(39,34){\line(2,3){9}}
\put(50,50){\line(-2,3){11}}
\put(50,50){\line(2,-3){11}}
\put(61,66){\line(1,-2){9}}
\put(70,48){\line(1,0){10}}
\put(30,48){\line(-1,0){10}}
\end{picture}
\begin{picture}(100,100) 
\put(50,50){\circle{40}}
\put(50,50){\line(2,3){11}}
\put(50,50){\line(-2,-3){11}}
\put(70,48){\line(1,0){10}}
\put(30,48){\line(-1,0){10}}
\put(61,67){\line(0,-1){34}}
\put(39,67){\line(0,-1){33}}
\end{picture}
\begin{picture}(100,100) 
\put(50,50){\circle{40}} 
\put(61,66){\line(-2,-3){9}}
\put(39,34){\line(2,3){9}}
\put(50,50){\line(-2,3){11}}
\put(50,50){\line(2,-3){11}}
\put(61,67){\line(0,-1){33}}
\put(70,48){\line(1,0){10}}
\put(30,48){\line(-1,0){10}}
\end{picture}
\vskip -5em 
\begin{picture}(100,100) 
\put(30,50){\line(-1,0){10}}
\put(70,50){\line(1,0){10}}
\put(50,50){\circle{40}}
\put(34,62){\line(1,0){32}}
\put(50,30){\line(-1,2){16}}
\put(50,30){\line(1,2){16}}
\end{picture}
\begin{picture}(100,100) 
\put(30,50){\line(-1,0){10}}
\put(50,50){\line(1,0){30}}
\put(50,50){\circle{40}}
\put(50,50){\line(-1,2){9}}
\put(50,50){\line(-1,-2){9}}
\put(41,32){\line(0,1){36}}
\end{picture}
\begin{picture}(100,100) 
\put(20,50){\line(1,0){60}}
\put(50,50){\circle{40}}
\put(50,30){\line(0,1){40}}
\end{picture}
\begin{picture}(100,100) 
\put(40,50){\circle{20}}
\put(66,50){\circle{30}}
\put(55,61){\line(1,-1){22}}
\put(77,61){\line(-1,-1){22}}
\put(30,50){\line(-1,0){10}}
\put(82,50){\line(1,0){10}}
\end{picture}
\vskip -5em 
\begin{picture}(100,100) 
\put(50,50){\circle{40}}
\put(20,50){\line(1,0){60}}
\put(42,68){\line(1,-1){18}}
\put(58,68){\line(-1,-1){18}}
\end{picture}
\begin{picture}(110,100) 
\put(20,50){\circle{20}}
\put(40,50){\circle{20}}
\put(60,50){\circle{20}}
\put(80,50){\circle{20}}
\put(10,50){\line(-1,0){10}}
\put(90,50){\line(1,0){10}}
\end{picture}
\begin{picture}(100,100) 
\put(19,50){\circle{20}}
\put(45,50){\circle{30}}
\put(55,50){\circle{30}}
\put(9,50){\line(-1,0){10}}
\put(71,50){\line(1,0){10}}
\end{picture}
\begin{picture}(100,100) 
\put(50,50){\circle{40}}
\put(50,60){\circle{20}}
\put(50,40){\circle{20}}
\put(30,50){\line(-1,0){10}}
\put(70,50){\line(1,0){10}}
\end{picture}
\vskip -5em 
\begin{picture}(100,100) 
\put(40,50){\circle{40}}
\put(60,50){\circle{40}}
\put(50,33){\line(-2,1){30}}
\put(20,48){\line(-1,0){10}}
\put(80,48){\line(1,0){10}}
\end{picture}
\begin{picture}(100,100) 
\put(50,50){\circle{40}}
\put(70,50){\line(1,0){10}}
\put(30,50){\line(-1,0){10}}
\put(50,30){\line(0,1){40}}
\put(30,50){\line(1,-1){20}}
\put(70,50){\line(-1,1){20}}
\end{picture}
\begin{picture}(100,100) 
\put(20,50){\circle{20}}
\put(40,50){\circle{20}}
\put(66,50){\circle{30}}
\put(10,50){\line(-1,0){10}}
\put(50,50){\line(1,0){42}}
\end{picture}
\begin{picture}(100,100) 
\put(40,50){\circle{20}}
\put(66,50){\circle{30}}
\put(66,66){\line(1,-1){16}}
\put(66,66){\line(-1,-1){16}}
\put(30,50){\line(-1,0){10}}
\put(82,50){\line(1,0){10}}
\end{picture}
\vskip -5em 
\begin{picture}(100,100) 
\put(50,50){\circle{40}}
\put(31,45){\line(-1,0){10}}
\put(69,45){\line(1,0){10}}
\put(38,66){\line(1,0){24}}
\put(38,66){\line(-1,-3){7}}
\put(62,66){\line(1,-3){7}}
\end{picture}
\begin{picture}(100,100) 
\put(50,50){\circle{40}}
\put(50,60){\circle{20}}
\put(20,50){\line(1,0){60}}
\end{picture}
\begin{picture}(100,100) 
\put(40,50){\circle{40}}
\put(60,50){\circle{40}}
\put(50,33){\line(0,1){34}}
\put(20,50){\line(-1,0){10}}
\put(80,50){\line(1,0){10}}
\end{picture}
\begin{picture}(100,100) 
\put(40,50){\circle{20}}
\put(60,53){\circle{20}}
\put(60,47){\circle{20}}
\put(30,50){\line(-1,0){10}}
\put(70,50){\line(1,0){10}}
\end{picture}
\vskip -5em 
\begin{picture}(100,100) 
\put(34,50){\circle{30}}
\put(66,50){\circle{30}}
\put(10,50){\line(1,0){80}}
\end{picture}
\begin{picture}(100,100) 
\put(50,50){\circle{40}}
\put(65,58){\circle{22}}
\put(58,68){\line(-3,-2){28}}
\put(30,49){\line(-1,0){10}}
\put(70,47){\line(1,0){10}}
\end{picture}
\begin{picture}(100,100) 
\put(50,50){\circle{40}}
\put(40,50){\circle{20}}
\put(60,50){\circle{20}}
\put(30,50){\line(-1,0){10}}
\put(70,50){\line(1,0){10}}
\end{picture}
\begin{picture}(100,100) 
\put(50,54){\circle{30}}
\put(50,46){\circle{30}}
\put(20,50){\line(1,0){60}}
\end{picture}

\section{Final remarks.}

The criterion (\ref{crit}) is easy to implement in the case 
when the polynomial $\tilde{P}$ in (\ref{1D}) depends only on a few variables
(large number of lines in the integral). In the most easy case when
number of lines is equal to number of "internal" scalar products
the irreducibility will follow from $P(0)\neq 0$.
In the case of a small number of lines (many variables in $\tilde{P}$)
it may be difficult to solve (\ref{crit}) explicitly. 
In this case it is instructive 
to check (using Groebner basis) if the polynomial $\tilde{P}$ is reducible to 
polynomials $\partial_{a \notin S} \tilde{P}$. If so, 
then $\partial_{a \notin S} \tilde{P}(\underline{x}^0)=0$ leads to 
$\tilde{P}(\underline{x}^0)=0$ and there no non--zero stable points. 
On the other hand, reducibility of the integrals with a small number of lines 
in many cases can be checked directly (with the triangle rule or by 
checking subdiagrams).

And finally, let us address the question what happens if (\ref{crit}) is not satisfied. 
Formally this case is unclear, but in practice (at least for massless propagator integrals
up to 4-loop and for zero-scale 2--loop propagator integrals \cite{Kalmykov}) it 
corresponds to reducible integrals.

The author are grateful to K.G.Chetyrkin, R.Harlander, J.H.Kuehn, Y.Schroder and 
O.V.Tarasov for 
the careful reading of the manuscripts and useful discussions.

\end{document}